\documentstyle[epsfig,psfig]{texas}

\def\spose#1{\hbox to 0pt{#1\hss}}
\def\gsim{\mathrel{\spose{\lower 3pt\hbox{$\mathchar"218$}}
          \raise 2.0pt\hbox{$\mathchar"13E$}}}
\def\lsim{\mathrel{\spose{\lower 3pt\hbox{$\mathchar"218$}}
          \raise 2.0pt\hbox{$\mathchar"13C$}}}

\begin{document}

\title{The origin of the relativistic wind in gamma-ray bursts:
MHD flow from a disk orbiting a stellar mass black hole?}

\author{F. Daigne and R. Mochkovitch}
\address{Institut d'Astrophysique de Paris\\
98bis Bd Arago, 75014 Paris, France\\
{\rm Email: daigne@iap.fr, mochko@iap.fr}}

%
% Abstract
%
\begin{abstract}
We compute the mass loss rate from a disk orbiting a stellar mass black hole 
assuming the flow is guided along magnetic field lines attached to the disk. 
We then estimate the Lorentz factor $\Gamma$ of the wind at infinity.
We find that $\Gamma$ can reach high values only if severe constraints 
on the field geometry and the conditions of energy injection are satisfied.
We discuss our results in the context of different scenarios for gamma-ray
bursts. We mention a risk of ``contamination'' of the Blandford-Znajek process
by wind material emitted from the disk. 

\end{abstract}

\section{Introduction}
Among the sources which have been proposed to explain cosmic gamma-ray
bursts (GRBs) the most popular are mergers of compact objects (neutron star
binaries or neutron star -- black hole systems) or
massive stars which collapse to a black hole (hypernovae) 
\cite{ref1, ref2, ref3}. In all 
cases, the resulting configuration is a stellar mass black hole surrounded
by a thick torus made of stellar debris or of infalling stellar material 
partially supported by centrifugal forces. 

If black hole + thick disk configurations are indeed at the origin of GRBs
the released
energy will ultimately come from the accretion of disk material by the black 
hole or from the rotational energy of the hole itself extracted by the
Blandford-Znajek mechanism. In a first step the energy must be injected into a
relativistic wind. The second step
consists in the conversion of a fraction of the wind kinetic energy
into gamma-rays via the formation of shocks, probably inside the wind itself
\cite{ref4, ref5}. 
In the last step 
the wind is decelerated when it interacts with the interstellar medium 
and the resulting (external) shock is responsible for the afterglow
observed in the X-ray, optical and radio bands \cite{ref6}.

The origin of the relativistic wind is certainly the more complex of the 
three steps. A few possible ideas have been proposed but none is presently 
fully conclusive. If the burst energy comes from matter accretion  
by the black hole, the annihilation of neutrino-antineutrino pairs 
emitted by the hot disk could be a way to inject energy along the system axis,
in a region which can be expected to be essentially baryon free due to the
effect of centrifugal forces. The low efficiency of this process however 
requires 
high neutrino luminosities and therefore short accretion time scales 
\cite{ref7}. 
Another possibility is to suppose that disk energy
is extracted by a magnetic field amplified by differential rotation to very 
large values ($B\sim 10^{15}$ G). A magnetically driven wind
could then be emitted from the disk with a fraction of the Poynting flux
being eventually transferred to matter. An alternative
to accretion energy could be to directly tap into the rotational energy of the
black hole via the Blandford-Znajek mechanism. The available power then depends
on the rotation parameter $a$ of the black hole and on the intensity of the
magnetic field pervading the horizon \cite{ref8}. 
The purpose of this paper is to present an exploratory study of the case where
a magnetically driven wind is emitted by the disk. Our approach will
be extremely simplified in comparison to the complexity of the real problem
so that our conclusions will have to be considered as indicative only. 
We nevertheless expect that we can identify the key parameters which control
the baryonic load of such a wind and put constraints on the final 
values of the Lorentz factor which can be obtained. 
\section{Dynamics of the wind from the disk to the sonic point}
To compute the mass loss rate and therefore estimate the amount 
of baryonic pollution we only need to follow the wind dynamics 
from the disk up to the sonic point. We write the wind equations with 
a number of simplifying assumptions: {\it i}) we assume that the disk 
is thin and
that the field is poloidal with the most simple geometry, i.e. straight lines
making an angle $\theta(r)$ with the plane of the disk, $r$ being the 
distance from the foot of the line to the disk axis (Fig. 1). The flow
of matter is then guided along the magnetic field lines; {\it ii}) we 
use non
relativistic equations since even at the sonic point $v_s/c<0.1$ but we adopt
the Paczynski-Wiita potential for the black hole; 
{\it iii}) we consider that a stationary regime has been reached in the 
wind.      
 
\begin{figure}
\centering
\epsfig{figure=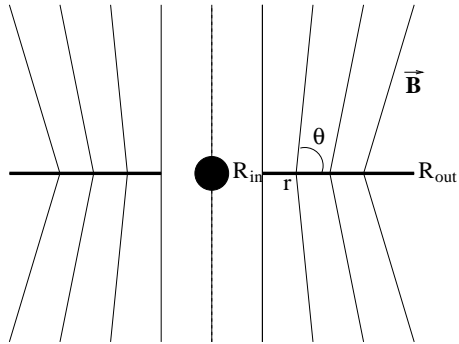, width=6.0cm}
\caption{Disk and field geometry}
\end{figure}

We then write the three flow equations in a frame corotating with the foot
of the line:
\begin{itemize}
\item Conservation of mass
\begin{equation}
\rho v s(y)={\dot m}
\end{equation}
\item Euler equation
\begin{equation}
v{dv\over dy}=\gamma(y)r-{1\over \rho}{dP\over dy}
\end{equation}
\item Energy equation
\begin{equation}
v {de\over dy}={\dot q}(y) r+v {P\over \rho^2}{d\rho\over dy}
\end{equation}
\end{itemize}
where $y=\ell/r$ and $\ell$ is the distance along the field line; 
$e$ is the specific internal energy, $\gamma(y)$ the total acceleration
(gravitational + centrifugal) and ${\dot q}(y)$ the power deposited in the
wind per unit mass. Different sources of heating can be present such as 
neutrino captures on nucleons (if the disk is optically thick to neutrinos),
neutrino-antineutrino annihilation and 
dissipation of kinetic or magnetic energy. Because the field and stream
lines are coincident the function $s(y)$ is easily related to the field 
geometry through the conservation of magnetic flux. Finally, ${\dot m}$ is
the mass loss rate from the disk per unit surface. Our equation of state 
which includes nucleons, relativistic electrons and positrons and photons 
is computed from the expressions given by \cite{ref9}.     

As long as the inclination angle $\theta$ remains larger than $\theta_1\simeq
60^{\circ}$ ($\theta_1$ is exactly $60^{\circ}$ if a newtonian instead
of a Paczynski-Wiita potential is used for the black hole) the acceleration 
$\gamma(y)$ is negative up to $y=y_1$ after which the centrifugal
force dominates. The sonic point of the flow is located just below $y_1$
(the relative difference $(y_1-y_s)/y_1$ never exceeds 1\%). 
To solve the wind equations 
we first fix trial values of the temperature and density $T_s$ and $\rho_s$ 
at the 
sonic point from which we get $v=v_s$. The position of the sonic point is 
obtained from the condition that the solution remains regular at $y=y_s$. 
The mass loss rate
${\dot m}$ is then fixed and the inward integration along a field line 
can be started. We
observe that at some position $y=y_{\rm crit}$ the velocity begins to fall off 
rapidly
while the temperature reaches a maximum $T_{\rm max}\le T_{\rm D}(r)$, where
$T_{\rm D}(r)$ is the disk temperature. We adjust the values    
of $T_s$ and $\rho_s$ with the requirement that $y_{\rm crit}$ should be as 
close as possible to 0 and $T_{\rm max}$ to $T_{\rm D}(r)$.

The results presented below assume that the disk is optically thick to neutrinos
between $R_{\rm in}=3\;r_g$ and $R_{\max}=10\;r_g$. No term for kinetic or 
magnetic energy dissipation have been included so that ${\dot q}(y)$ is limited
to neutrino processes: capture
on free nucleons, scattering on electrons and positrons, neutrino-antineutrino
annihilation (heating) and neutrino emission by nucleons, annihilation of
electron-positron pairs (cooling). The assumption that the disk is optically 
thick to neutrinos is probably justified for NS + NS or
NS + BH mergers. It is much more questionable in the hypernova scenario,
except for very high accretion rate or low values of
the viscosity parameter ($\alpha< 0.01$) 
as shown from the disk models computed by \cite{ref10}. Our detailed 
results 
therefore only concern a specific case but we also     
obtain below a simple and general rough estimate of the mass loss rate for any 
kind of heating mechanism. 

The adopted temperature distribution $T_{\rm D}(r)$ corresponds to a geometrically
thin, optically thick disk
\begin{equation}
T_{\rm D}(r)=T_*\left({r_*\over r}\right)^{3/4}\left(
{1-\sqrt{r_{\rm in}\over r}\over {1-\sqrt{r_{\rm in}\over r_*}}}\right)^{1/4}
\end{equation}
where $T_*$ is the temperature at $r=r_*$. The mass of the black hole is
$M_{\rm BH}=2.5$ M$_\odot$.  
\section{The mass loss rate}
The solution for mass loss rate as a function of $r$, $T_{\rm D}(r)$ and
$\theta(r)$ takes the form
\begin{equation}    
\eqalign{
&{\dot m}_{13}(x)\approx 3.8\;\mu_{\rm BH}\left[{T_{\rm D}(x)\over 2\ {\rm MeV}}
\right]^{10}f[x, \theta(x)]\cr
&\approx 3.8\;\mu_{\rm BH}\left[{T_*\over 2\ {\rm MeV}}\right]^{10}
\left({r_*\over r}\right)^{15/2}\left(
{1-\sqrt{r_{\rm in}\over r}\over {1-\sqrt{r_{\rm in}\over r_*}}}\right)^{5/2}
f[x, \theta(x)]}
\end{equation}
where ${\dot m}_{13}$ is the mass loss rate in units of $10^{13}$ 
g.cm$^{-2}$.s$^{-1}$, $x=r/r_g$ and $\mu_{\rm BH}=M_{\rm BH}/2.5\;M_\odot$.
The geometrical function $f[x,\theta(x)]$ is normalized in such a way that it
is equal to unity for $x=4$ and
$\theta=85^{\circ}$.
The mass loss rate is extremely sensitive to the value of the disk temperature.
The tenth power dependence is in agreement with what is found for neutrino
driven winds in spherical geometry \cite{ref11}. 
The dependence of ${\dot m}$ on
inclination angle is also very strong as shown 
in Fig. 2 where ${\dot m}$ is represented (with $T_*=2$ MeV and 
$r_*=4\;r_g$) for two geometries of the field lines: constant 
$\theta=85^{\circ}$ and $\theta$ decreasing from $90^{\circ}$ to 
$80^{\circ}$ between $r=3\;r_g$ and $r=10\;r_g$.    

\begin{figure}
\centering
\epsfig{figure=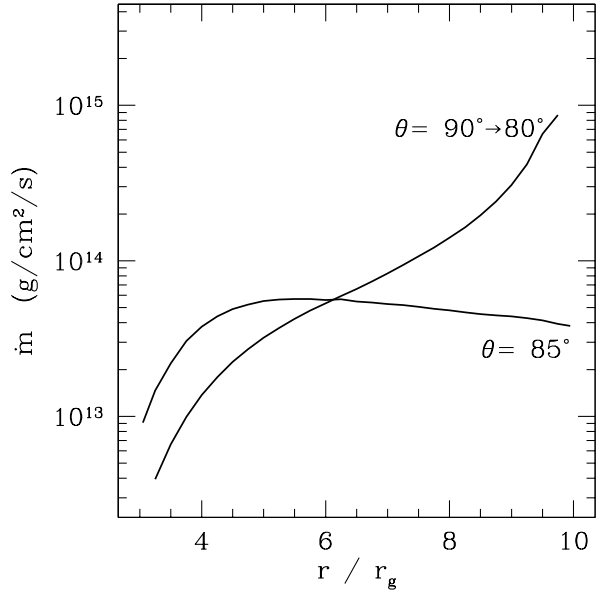, width=8.0cm}
\caption{Mass loss rate from the disk for a constant and 
decreasing inclination angle of the field lines. The disk temperature is 2 MeV
at $r=4\,r_g$}
\end{figure}

Since additional sources of heating can be present in the wind (viscous 
dissipation,  
reconnection of field lines, etc) we have also obtained a very simple and 
general
analytical expression for the mass loss rate \cite{ref12} 
\begin{equation}
{\dot m}\approx {{\dot e}\over \Delta \Phi}g
\end{equation}
where ${\dot e}$ is the rate of thermal energy deposition (in erg.cm$^{-2}$.s$^{-1}$)
between the plane of the disk ($y=0$) and the sonic point at $y_s\simeq  y_1$; 
$\Delta\Phi=\Phi_1-\Phi_0$ is the difference of potential (gravitational +
centrifugal) between $y=0$ and $y=y_1$. The $g$ factor, which is of the order 
of unity, depends on the distribution of energy injection between  
$y=0$ and $y=y_1$.
\section{Average Lorentz factor of the wind}
To estimate the Lorentz factor which can be reached by the wind one must
be able to relate the injected energy to the mass loss rate. This can be 
done in the following way: we suppose that we observe a burst power in
gamma-rays
\begin{equation}  
{\dot {\cal E}}_{\gamma}={10^{51}\over 4 \pi} \,\epsilon_{51}
\ \ \ {\rm erg.s}^{-1}.{\rm sr}^{-1}
\end{equation}
Then, the power injected into the wind was
\begin{equation}
{\dot E}=2\;10^{51} {f_{\Omega}^{0.1}\over f_{\gamma}^{0.05}}\; \epsilon_{51}
\ \ \ {\rm erg.s}^{-1}
\end{equation}
where $f_{\Omega}^{0.1}$ and $f_{\gamma}^{0.05}$ are respectively the fraction ${\Delta\Omega \over 4 \pi}$ of solid angle covered by the wind 
(in unit of 0.1) and the efficiency for the conversion of kinetic energy 
into gamma-rays (in unit of 0.05).  
Accretion by the black hole powers the wind but at the same time 
viscous dissipation heats the disk which cools by the emission of neutrinos.
If neutrino losses represent a fraction $\alpha$  of the energy ${\dot E}$
injected into the wind we have (for an optically thick disk) 
\begin{equation}
{\dot E}_{\nu}=\alpha {\dot E}=2\;10^{51} 
{f_{\Omega}^{0.1}\over f_{\gamma}^{0.05}}\; \alpha \epsilon_{51}=
2\int_{r_{\rm in}}^{r_{\rm out}}{7\over 8}\sigma T_{\rm D}^4(r)\;2\pi r dr
\end{equation}
Substituting in (9) Eq. (4) for $T_{\rm D}(r)$ we obtain for the temperature
at $r_*=4 \;r_g$
\begin{equation}
T_*=1.72\,\mu_{\rm BH}^{-1/2}\left(
{f_{\Omega}^{0.1}\over f_{\gamma}^{0.05}}\; \alpha \epsilon_{51}\right)^{1/4}
\ \ \ {\rm MeV}
\end{equation}
The value of $T_*$ being known ${\dot m}$ can be computed as a function of 
of $r$ for a given field geometry. The total mass loss rate from the disk
is then 
\begin{equation}
\eqalign{
{\dot M}&=2\int_{r_{\rm in}}^{r_{\rm out}} {\dot m}(r)\, 2\pi r dr\cr
&=2.6\;10^{26} \mu_{\rm BH}^3\left({T_*\over 2\ {\rm MeV}}\right)^{10} \;
{\cal F}}
\end{equation}
where 
\begin{equation}
{\cal F}=\int_{r_{\rm in}/r_g}^{r_{\rm out}/r_g} 
f[x, \theta(x)]\, x dx
\end{equation}
is a function of the field geometry.
The average Lorentz factor is finally given by 
\begin{equation}
{\bar \Gamma}={{\dot E}\over {\dot M}c^2}=
{8500\over {\cal F}} \mu_{\rm BH}^2\, \epsilon_{51}^{-3/2}\,
\alpha^{-5/2}\left({f_{\gamma}^{0.05}\over f_{\Omega}^{0.1}}\right)^{3/2}
\end{equation}
The value of ${\cal F}$ is 56 for a constant inclination angle 
$\theta=85^{\circ}$ and 250 if
$\theta$ decreases from $90^{\circ}$ to $80^{\circ}$ between $r=3$ and $10\ r_g$. These numbers shows
that large Lorentz factors can be reached but only under quite
restrictive conditions:  
quasi vertical field lines, low $\alpha$ values, i.e. good efficiency for 
energy injection into the wind with little dissipation and necessity of beaming.

More generally, Eq. (6) can also provide a simple and useful constraint on
the terminal Lorentz factor. If the power ${\dot e}$ deposited
below the sonic point represents
a fraction $x$ of the total power ${\dot e}_{\rm tot}$ which is finally 
injected into the wind we get 
\begin{equation}
\Gamma\approx {{\dot e}_{\rm tot}\over {\dot m}c^2}\approx
{\Delta\Phi/c^2\over g x}
\end{equation}
Considering a line anchored at $r=4\,r_g$ with an inclination angle  
$\theta=85^{\circ}$ we obtain $y_1 = 2.182$ and $\Delta\Phi/c^2=0.18$ which implies that $x$ 
should not exceed $10^{-3}$ to have $\Gamma>100$! 
This is clearly a very strong constraint on any mechanism of energy injection.
The wind however remains relativistic for $x\lsim 0.1$ but its Lorentz factor
is then much too low to produce a cosmic GRB.
\section{Discussion}
We have computed the mass loss rate in a magnetically driven wind emitted 
by a disk orbiting a stellar mass black hole. Detailed results are given
for the case where the disk is optically thick to neutrinos but we have
also obtained an approximate analytical expression, valid for any heating 
mechanism. From the mass loss rate the terminal Lorentz factor of the wind
can be estimated. Large values of $\Gamma$ require that severe constraints 
on the field geometry and the amount of heating below the sonic point should
be satisfied. Another potential problem which was not addressed in the present 
paper is
that relativistic MHD winds, at least in their simplest version \cite{ref13},
can be quite inefficient in transferring magnetic into kinetic energy. 

An optimistic view of the situation would be to consider that the
difficulty to obtain high Lorentz factors could just be a way to explain
the apparent discrepancy between the birthrate of the sources in the hypernova
scenario $\sim 10^{-3}$ yr$^{-1}$/galaxy (for very massive stars) and the
observed GRB rate $\lsim 10^{-6}$ yr$^{-1}$/galaxy. Beaming alone
cannot account for the difference which implies that the collapse
of a massive star most generally fails to give a GRB. 
 
The pessimistic view naturally consists to conclude that the baryonic load
of the wind emitted by the disk is so large that the Lorentz factor can never 
reach values of $10^2$ and more. The next step is then to rely 
on the Blandford-Znajek
mechanism to produce the relativistic wind \cite{ref8}. One should be careful 
however that magnetic field lines coming from the disk and trapped by the
black hole will also carry frozen in wind material  
leading to a ``contamination'' of the Blandford-Znajek process by the disk. 
A possible loophole could be that the accretion time scale to the black hole
is so short that the stationary wind solutions we have obtained are not valid.
The amount of material extracted from the disk could therefore be much smaller 
still allowing a highly relativistic outflow to develop.

\end{document}